\documentclass[preprint,proceedings]{rmaa}

\newcommand{\D}{\discretionary{}{}{}}

\SetYear{2006}
\SetConfTitle{First Light Science with the GTC}

\title{New Results from Observations of Massive Star Formation in the Mid-Infrared with Large Aperture Telescopes}

\author{James M. De Buizer \altaffilmark{1}}

\altaffiltext{1}{Gemini Observatory, Chile (jdebuizer\D{}@gemini.\D{}edu.)}

\suppressfulladdresses

\listofauthors{J. M. De Buizer}
\indexauthor{De Buizer, J. M.}

\addkeyword{Circumstellar matter}
\addkeyword{ISM: Jets and outflows}
\addkeyword{Stars: Formation}
\addkeyword{Stars: Early type}

\begin{document}
% Typeset article header
\maketitle

\boldabstract{Thanks to the high spatial resolution afforded by
8-10m class telescopes, we are beginning to learn that some sources
are extended in their mid-infrared emission because of dusty
outflows or heated outflow cavity walls. Therefore one must be
extremely careful in interpreting the nature of extended
mid-infrared sources (i.e. just because it is extended does not
automatically mean it is a disk!).}

Recent high spatial resolution observations of massive young stellar
sources have yielded evidence of MIR emission (5-25 $\micron$) from
outflows and jets. These observations correlate well with other
larger-scale outflow indicators and their geometries, such as what
is seen in shock-excited H$_2$ and CO emission.

De Buizer \& Minier (2005) and De Buizer (2006a) showed two massive
young stellar sources with clear MIR signatures from outflow.
Another source that might be added to this list is IRAS 20126+4104
(De Buizer 2006b). This source is considered THE prototype of a
young massive star with a circumstellar disk (see Cesaroni et al.
2006 for a review). The ``disk'' can be seen in mm continuum as well
as molecular line emission, but because it is thousands of AU in
size, there is disagreement as to whether this ``disk'' is a real
circumstellar accretion-type disk or a larger flattened envelope
that feeds a smaller central accretion disk. There is also a
large-scale outflow at an angle close to perpendicular to the
``disk''.

\begin{figure}[!t]
  \includegraphics[width=\columnwidth]{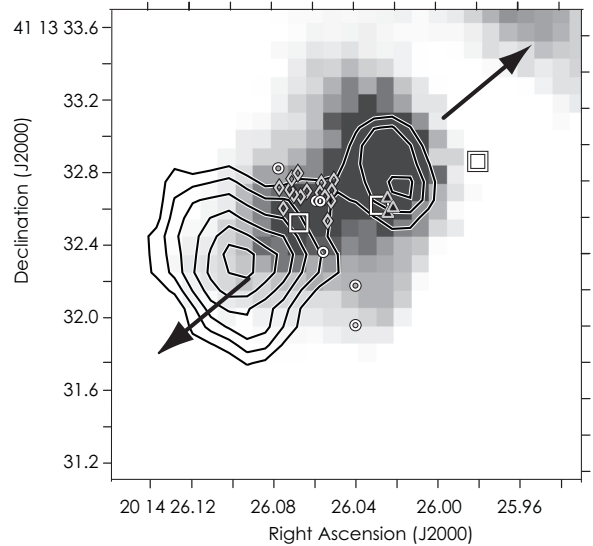}
  \caption{The deconvolved 12.5 $\micron$ image of IRAS 20126+4104.
  The black contours are the 2 $\micron$ emission of Sridharan et al.
  (2005). Maser components are shown: OH (circles), water (triangles),
  and methanol (diamonds) from Edris et al. (2005); and water (squares)
  from Tofani et al. (1995). Arrows show the outflow angle.}
  \label{fig:simple}
\end{figure}

Recently, the 2 $\micron$ images of Sridharan et al. (2005) showed a
double-lobed structure separated by a ``dark lane'' here at the
center of IRAS 20126+4104. It was claimed that this was a silhouette
disk, similar to those seen around low mass stars. It was also
claimed that the 5 $\micron$ emission may be coming from the thermal
dust emission from the central accretion disk in the dark lane.
Using Michelle on Gemini North high spatial resolution MIR images of
this region were obtained. It was found that the thermal dust
emission is distributed in double-lobed structure with a dark lane
at the same location of the dark lane in the 2 $\micron$ emission.
The 5 $\micron$ emission was found to be coincident with the
southeastern MIR emission lobe and therefore no thermal emission is
actually seen directly from the dark lane. Figure 1 shows that the
MIR and near-IR emission are likely coming from the outflow cavities
centered on some unseen source at the center of the dark lane. The
maser emission in this region appears to be associated with the IR
emission of the outflow cavities and not the dark lane itself.

\end{document}